\documentclass[reprint,twocolumn,floatfix,aps]{revtex4}
\usepackage{CJK}
\usepackage{amsmath}%
\usepackage{graphicx}%
\usepackage{dcolumn}%
\usepackage{physics}%
\usepackage{bm}%
\usepackage{epsfig,colordvi}
\usepackage{color}
\usepackage[normalem]{ulem}
\usepackage{soul}

\usepackage[normalem]{ulem}

\begin{document}

\title{Cluster formation near midrapidity - can the mechanism be identified experimentally?}

\author{V.~Kireyeu$^{1,2}$,
G.~Coci$^{1,3}$, S.~Gl{\"a}{\ss}el$^{4}$,  J.~Aichelin$^{5,6}$, C.~Blume$^{4}$, E.~Bratkovskaya$^{7,3,1}$}

\affiliation{$^{1}$ Helmholtz Research Academy Hessen for FAIR (HFHF),GSI Helmholtz Center for Heavy Ion Physics. Campus Frankfurt, 60438 Frankfurt, Germany}
\affiliation{$^{2}$ Joint Institute for Nuclear Research, Joliot-Curie 6, 141980 Dubna, Moscow region, Russia}
\affiliation{$^{3}$ Institut f\"ur Theoretische Physik, Johann Wolfgang Goethe University,
Max-von-Laue-Str. 1, 60438 Frankfurt, Germany}
\affiliation{$^{4}$ Institut f\"ur Kernphysik, Max-von-Laue-Str. 1, 60438 Frankfurt, Germany}
\affiliation{$^{5}$ SUBATECH, Nantes University, IMT Atlantique, IN2P3/CNRS
4 rue Alfred Kastler, 44307 Nantes cedex 3, France}
\affiliation{$^{6}$ Frankfurt Institute for Advanced Studies, Ruth Moufang Str. 1, 60438 Frankfurt, Germany} 
\affiliation{$^{7}$ GSI Helmholtzzentrum f\"ur Schwerionenforschung GmbH,
  Planckstr. 1, 64291 Darmstadt, Germany}
  
\date{\today}

\begin{abstract} \noindent
The formation of weakly bound clusters in the hot and dense environment at midrapidity is one of the surprising phenomena observed experimentally in heavy-ion collisions from a low center of mass energy of a few GeV up to a ultra-relativistic energy of several TeV. Three approaches have been advanced to describe the cluster formation: coalescence at  kinetic freeze-out, cluster formation during the entire heavy-ion collision by potential interaction between nucleons and deuteron production by hadronic kinetic reactions. Based on the Parton-Hadron-Quantum Molecular Dynamics (PHQMD) microscopic transport approach, which incorporates all 3 mechanisms for deuteron production, we identify experimental observables, which can discriminate these production mechanisms for deuterons.
\end{abstract}


\maketitle

\section{Introduction}
The observation of light clusters at midrapidity  \cite{ALICE:2015wav,ALICE:2017nuf} was one of the biggest surprises of the heavy ion experiments at ultra-relativistic energies. At midrapidity the transverse energy spectra of all hadrons have slopes of more than 100 MeV and point towards a very hot interaction region. Simulations of these reactions by transport and hydrodynamic models predict energy densities of well above 1 $\rm GeV/fm^3$ and hence a very dense interaction zone. In such an environment it is difficult to understand how loosely bound objects like $d$ and $t$, with binding energies in the one MeV region, can survive. The thermal momentum of possible interaction partners would be sufficient to destroy them.  This observation has been named ``ice in the fire puzzle".

It was even more surprising that in ultra-relativistic heavy-ion collisions the multiplicity of these clusters follows, without exception, the prediction of the statistical model if the same value of the temperature as for all the other hadrons is employed~\cite{Andronic:2010qu}. The statistical model predicts the multiplicity of hadrons at chemical freeze-out and hence before the final state interactions among the hadrons which continue until the kinetic freeze-out. Taken for granted that the prediction of the cluster multiplicity by statistical model calculations is not a pure accident, one has to understand how it is possible that clusters survive the hadronic expansion from chemical to kinetic freeze-out.

\section{kinetic models for deuteron production}

The observation of  these light clusters has also renewed the general theoretical interest in cluster production in heavy ion reactions. Besides the statistical model three approaches have been advanced and applied for the cluster production within the kinetic transport approaches:

$\bullet$ {\bf  Coalescence mechanism} 
(cf. \cite{Butler:1963pp, Sun:2018jhg, Kittiratpattana:2022knq}
and reference therein). \\
The coalescence mechanism is mostly used for deuteron production and assumes that clusters are formed at kinetic freeze-out, i.e. after the hadronic expansion phase, when the last of the two constituents of the deuteron had its last hadronic collisions. 
If at that time a fellow nucleon with the right charge is in the coalescence radius in coordinate and momentum space, the two nucleons are considered as a deuteron. The radii are determined by a fit of the experimental multiplicities.  Another variance of the coalescence model uses the Wigner density of the deuteron to determine the coalescence probability  \cite{Scheibl:1998tk,Zhu:2015voa}. The model can be extended to larger clusters, however the number of parameters, necessary  to determine larger clusters, increases fast. \\
$\bullet$  {\bf Cluster by potential interactions among nucleons - 'potential' mechanism} \cite{Aichelin:1991xy,Aichelin:2019tnk,Glassel:2021rod}.\\
The potential interaction of nucleons during the hadronic phase creates bound clusters of any size whose multiplicity depends on the details of the expansion of the hot interaction zone and its composition. This dynamical formation of clusters by potential interactions of nucleons can be modeled by propagating the n-body phase space density  as done in the quantum molecular dynamics (QMD) approach. 
Clusters are not suddenly produced but identified during the dynamical evolution of the system by the  Minimum Spanning Tree (MST) procedure. The MST combines nucleons to clusters using a closest distance algorithm, i.e. a nucleon is a part of a cluster if its distance to the closest nucleon is smaller than a radius $r_{clus}$ which is about the range of the nucleon-nucleon interaction.
The implementation of this procedure in the PHQMD microscopic transport approach is explained in Ref. \cite{Aichelin:2019tnk}.
The MST procedure has been advanced further to aMST ("advanced MST") in Ref. \cite{Coci:2023daq} by selecting only clusters with negative binding energies and by introducing  a 'stabilization procedure', which allows to eliminate the artificial emission of nucleons of bound clusters
due to the semi-classical QMD dynamics. 
We stress that MST is a cluster recognition procedure, not a 'cluster building' mechanism, since the QMD transport approach propagates baryons, not clusters. \\
$\bullet$ {\bf Cluster production by collisions - 'kinetic' mechanism}~\cite{Oliinychenko:2018ugs,Oliinychenko:2020znl,Staudenmaier:2021lrg,Coci:2023daq} \\ Light clusters - as deuterons - can be produced by inelastic reactions of hadrons like $NN\pi \to d\pi$ and $NNN\to dN$ with a pion or a nucleon as a ``catalyst" during hadronic phase of heavy-ion collisions. They are called `kinetic' deuterons.  
 In the first application of this `kinetic' mechanism in the transport model SMASH, the three-body $3\to 2$ reactions have been replaced by two two-body collisions with an intermediate  fictitious dibaryon resonance $d^*$ \cite{Oliinychenko:2018ugs}. In the meantime the three-body entrance channel has been modeled directly employing detailed balance to the experimentally measured \cite{Staudenmaier:2021lrg} inverse $\pi d$ and $N d$ scattering.
 
In our recent work \cite{Coci:2023daq} we extend the study of Ref. \cite{Staudenmaier:2021lrg} by implementing all isospin channels for pion induced reactions in the microscopic Parton-Hadron-Quantum Molecular Dynamics approach (PHQMD) \cite{Aichelin:2019tnk} as well as deuteron finite size effects. In the `kinetic' approach deuteron production and destruction is possible during the whole hadronic expansion of the hot interaction zone but calculations show that only the later produced cluster survive.

In all three approaches the transverse momentum distribution of clusters has been calculated 
(cf. \cite{Oliinychenko:2018ugs,Staudenmaier:2021lrg,Kireyeu:2022qmv,Glassel:2021rod,Coci:2023daq}).
The comparison with the experimental spectra in a wide energy range - from center-of-mass energies between 2.5 GeV  to  several TeV - shows for all approaches a reasonable agreement despite of the fact that deuterons are produced very differently:  i) at different times: in the coalescence model deuterons are produced at freeze-out while in the other models they are created by interactions during  the hadronic expansion phase; 
ii) according to different criteria:  by phase-space correlations of nucleons at freeze-out in the coalescence scenario, by space correlations of nucleons  with the condition that they are bound  in the potential scenario, where in both approaches the deuteron is an extended object, by hadronic collisions where the  deuterons are approximated as point like particles.

\section{Results within the PHQMD}

This situation that several theoretical models for cluster production, based available on different assumptions, describe the same experimental data (in the  experimentally available rapidity and $p_T$ range, which often doesn't cover the low $p_T$ domain) is enigmatic and unsatisfying. To understand the production of these clusters it is, therefore, desirable to explore whether there are experimental observables, which can discriminate between the different approaches.  This is the goal of this study. 

To make this possible we have developed the PHQMD approach  \cite{Aichelin:2019tnk,Glassel:2021rod,Kireyeu:2021igi,Kireyeu:2022qmv,Coci:2023daq} further. It allows now to realize all three mechanisms for deuteron production, `coalescence' deuterons, `potential' deuterons and 'kinetic' deuterons, in the same transport approach. This allows to study the three different production mechanisms in an otherwise identical environment. 

We recall that in PHQMD the 'potential' deuterons are identified by the aMST cluster finding algorithm during the whole time evolution as described in Refs. \cite{Aichelin:2019tnk,Coci:2023daq} while the 'kinetic' deuterons are produced  by inelastic hadronic collisions - cf. Ref. \cite{Coci:2023daq} for the details. 
The coalescence mechanism in PHQMD \cite{Kireyeu:2022qmv} is adopted from the UrQMD model \cite{Botvina:2014lga,Sombun:2018yqh}: the deuterons are formed at freeze-out if the 
 relative momentum $\Delta P$ and distance $\Delta R$ between the proton and the neutron in its center-mass frame  $\Delta P<0.285$ GeV and $\Delta R< 3.575$ fm. 
We have made sure that PHQMD in the coalescence option agrees with the previous PHQMD coalescence calculations \cite{Kireyeu:2022qmv} - where we compared the coalescence and MST mechanisms implemented in  the PHQMD and UrQMD transport approaches.

\begin{figure}[t!]
\centering
\includegraphics[width=0.45\textwidth]{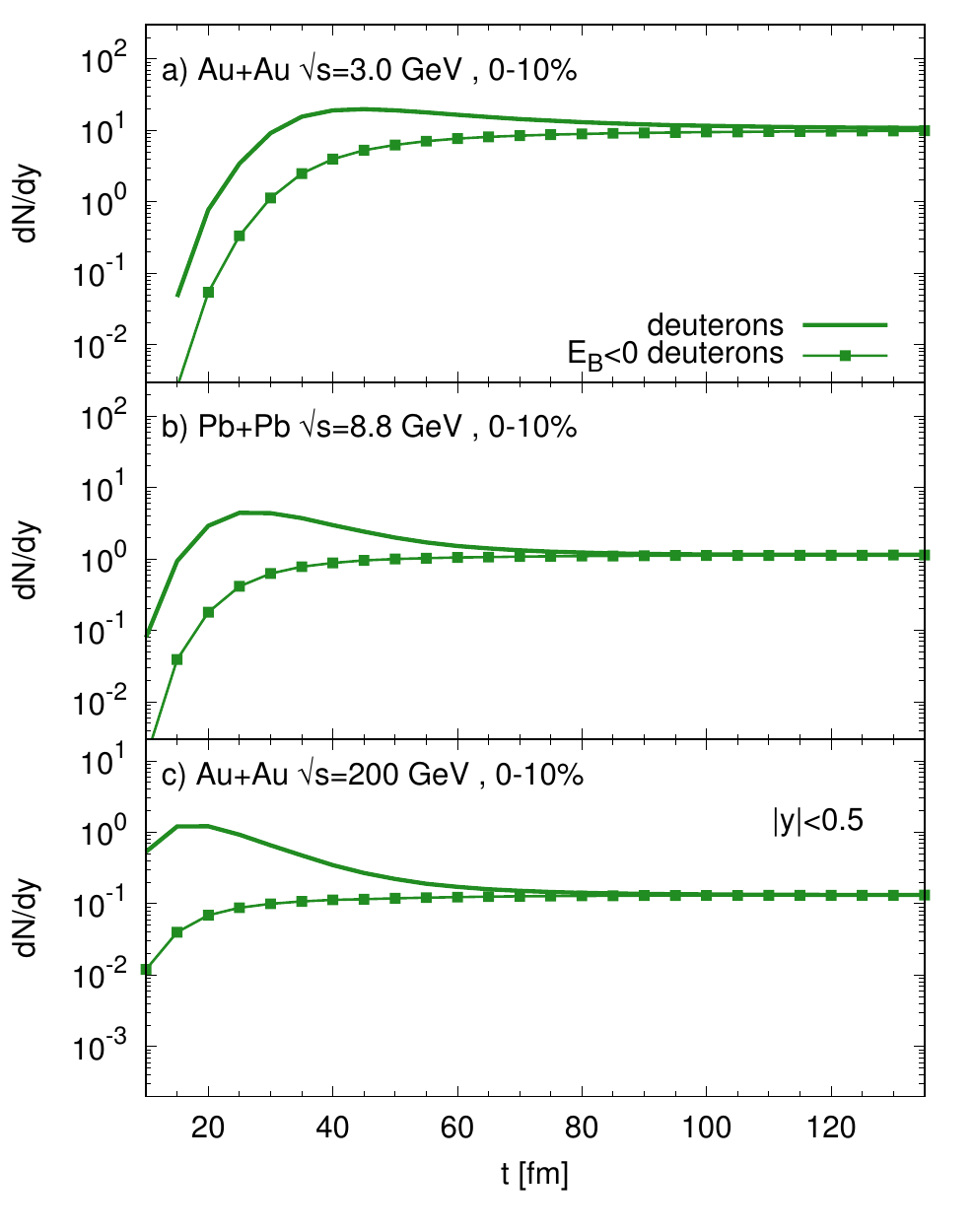}
\label{multtime}
\caption{\label{fig:AS} Multiplicity of deuterons in central collisions  at mid-rapidity, $|y|<0.5$ in PHQMD simulations for  Au+Au at $\sqrt{s}_{NN}=3.0$ GeV (top), for Pb+Pb at $\sqrt{s}_{NN}=8.8$ GeV (middle),  and for Au+Au at $\sqrt{s}_{NN}=200$ GeV (bottom). The solid lines are the results of MST, the solid lines with filled squares show the bound clusters, ($E_B<0$ ) analyzed with aMST, whose difference to MST is explained in Ref. \cite{Coci:2023daq}. } 
\label{Fig:fig1}
\end{figure}

We start out with the time evolution of the clusters. The time evolution of the midrapidity multiplicity of bound deuterons and of all `deuterons' (bound and unbound) in the `potential' production mechanism is shown in Fig. \ref{fig:AS} for central Au+Au collisions at three different beam energies.  The deuterons are identified by MST as two nucleons with a distance $r\le 4 \, {\rm fm}$, the range of the $NN$ potential, under the condition that no other baryon has a distance smaller than 4 fm to one of the two nucleons.  As seen in Fig. \ref{fig:AS}, the total number of deuterons (solid line) - identified by MST - grows with time and then decreases because only bound clusters survive asymptotically (the solid line with filled squares) \cite{Glassel:2021rod}. The nucleons in unbound deuterons separate due to their momentum difference and it is just a question of time when their distance gets larger than 4 fm, the limit that the nucleons are considered as a cluster. 
Freeze-out takes place around 25 fm/c at $\sqrt{s}$ = 2.52 GeV and decreases slightly with energy \cite{Glassel:2021rod}. Roughly at that time it is determined in the coalescence model whether two nucleons form a deuteron. We see here that at this time the number of MST `deuterons'  (solid line) is larger than for $t\to \infty$. Most of the MST `deuterons' at freeze-out are `unbound' and the nucleons separate from each other during the further propagation.

\begin{figure}[t]
    \centering
        \includegraphics[width=0.45\textwidth]{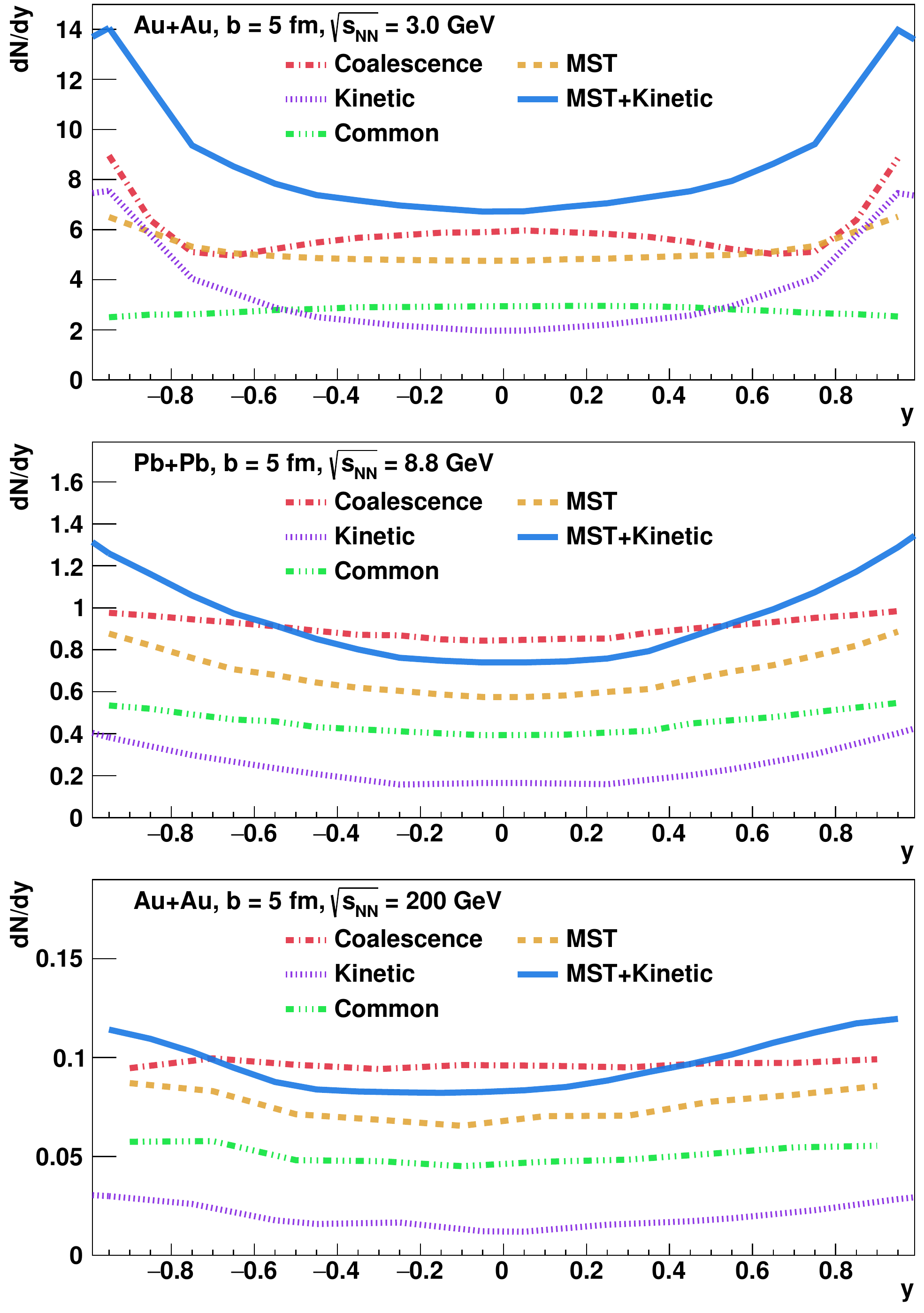} 
    \caption{Rapidity distribution of deuterons in central collisions at a center-of-mass energy  of  $\sqrt{s}_{NN}=3$ GeV (top), $\sqrt{s}_{NN}=8.8$ GeV (middle), and  $\sqrt{s}_{NN}=200$ GeV (bottom). 
    The orange dashed lines show the results of the aMST with $E_B<0$ for `potential' deuterons, the dash-dotted red lines  that for coalescence deuterons and the dotted purple lines that for `kinetic' deuterons produced by collisions. The solid blue lines are for the sum of the aMST and kinetic deuterons. The green dash triple-dot lines indicate the rapidity distribution of those `potential' deuterons, identified by the aMST, which are at the same time `coalescence' deuterons without applying the factor 3/8.}
    \label{fig:dndy}
\end{figure}

\begin{figure}[t]
    \centering
        \includegraphics[width=0.45\textwidth]{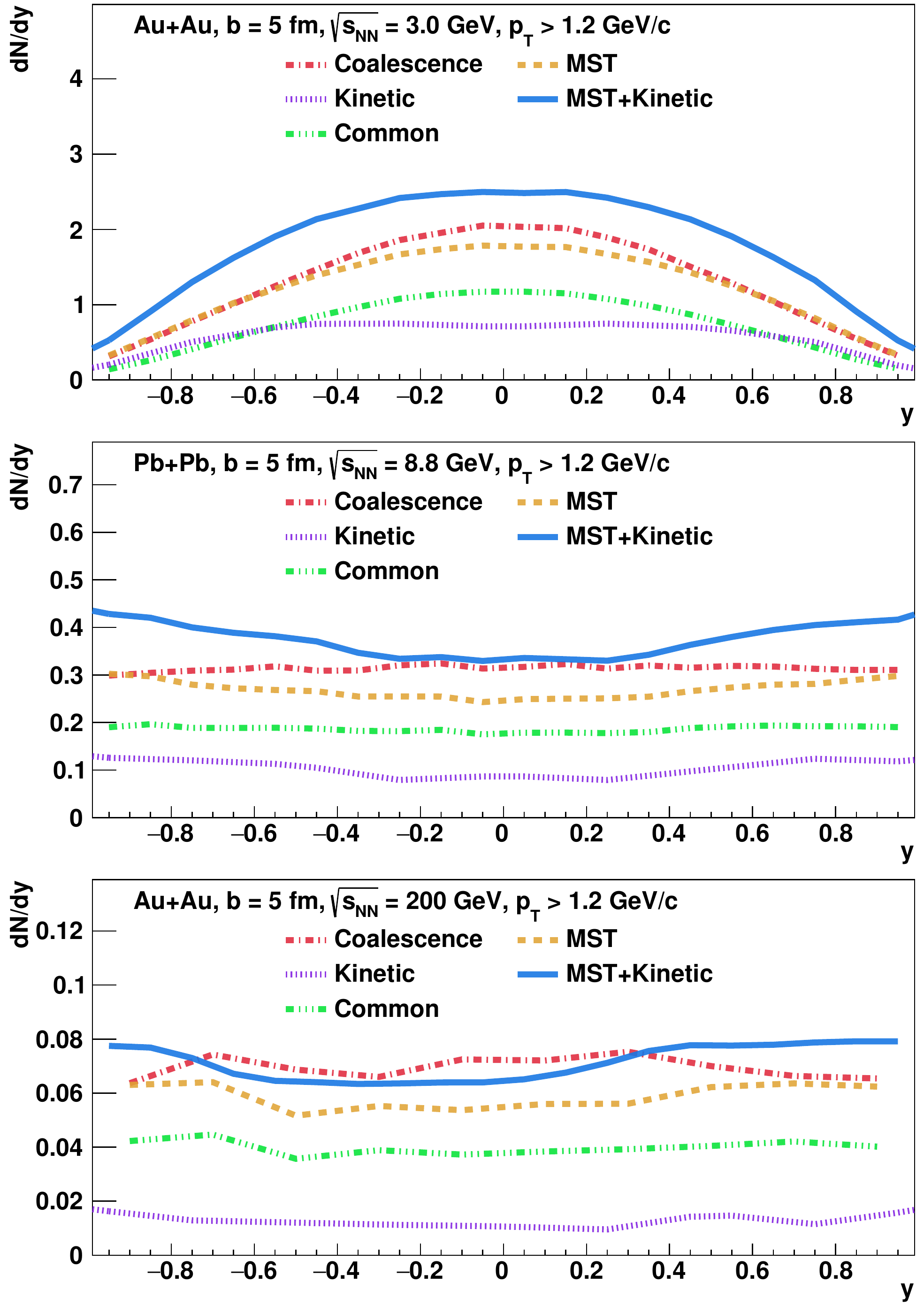} 
    \caption{
    Same as Fig.\ref{fig:dndy}  for deuterons with  $p_{T} > 1.2$ GeV/c. }
    
    \label{fig:dndy_ptcut}
\end{figure}

The rapidity distributions of deuterons are presented in Fig. \ref{fig:dndy} for the same center-of-mass energies as the multiplicities in Fig. \ref{fig:AS}.  The orange dashed lines show the results for potential deuterons with $E_B<0$ (aMST), the dash-dotted red lines stand for the deuterons found in the coalescence approach. For the coalescence radii we used the values of Ref. \cite{Kireyeu:2022qmv}:  $|r_1-r_2| \le 3.575$ fm and $|p_1-p_2|\le 285$ MeV. These radii have been fitted to data in order to reproduce the deuteron multiplicity if a spin degeneracy factor of 3/8 \cite{Kireyeu:2022qmv} is applied. The multiplicity of `coalescence' deuterons, which are formed at freeze-out, is much larger than that of  the final (bound) `potential' deuterons. This can been inferred from Fig. \ref{Fig:fig1}, which shows that the number of all `potential' deuterons at that time is larger than the asymptotic yield. Therefore, if we apply the PHQMD nucleon-nucleon potential most of the `coalescence' deuterons are unbound. The lines for `coalescence' deuterons, presented in Fig. \ref{fig:dndy}, show the number of nucleon pairs, which fulfil the coalescence criteria, multiplied by the factor 3/8. 
Furthermore, the dotted purple lines  present the rapidity distribution of the `kinetic' deuterons, which are much suppressed in PHQMD by including  finite-size effects  in coordinate and momentum space by two conditions - an excluded volume (no surrounding particle in the radius of deuterons) and  momentum projection of nucleons  forming a deuteron onto the deuteron wave function \cite{Coci:2023daq}. The solid blue lines represents the sum of the  potential deuterons with $E_B<0$ and `kinetic' deuterons.
Finally, the green dash triple-dot lines indicate the rapidity distribution of `potential' deuterons, identified by the aMST, which are simultaneously  `coalescence' deuterons. 
We observe that at midrapidity only about 20\% of  the MST deuterons are also identified as deuterons in the coalescence approach - before multiplication of coalescence results by the factor 3/8.
Because potential deuterons and coalescence deuterons are composed of different nucleons  we may expect to see differences in the observables. Indeed, the rapidity distribution of deuterons is rather different for the different approaches.  At low energy the rapidity distribution of coalescence deuterons is concave and shows a maximum at midrapidity, which flattens out for higher energies, whereas potential deuterons have a rather flat distribution at low energy developing a convex form around midrapidity at higher energies.

\begin{figure}[t]
\centering
        \includegraphics[width=0.45\textwidth]{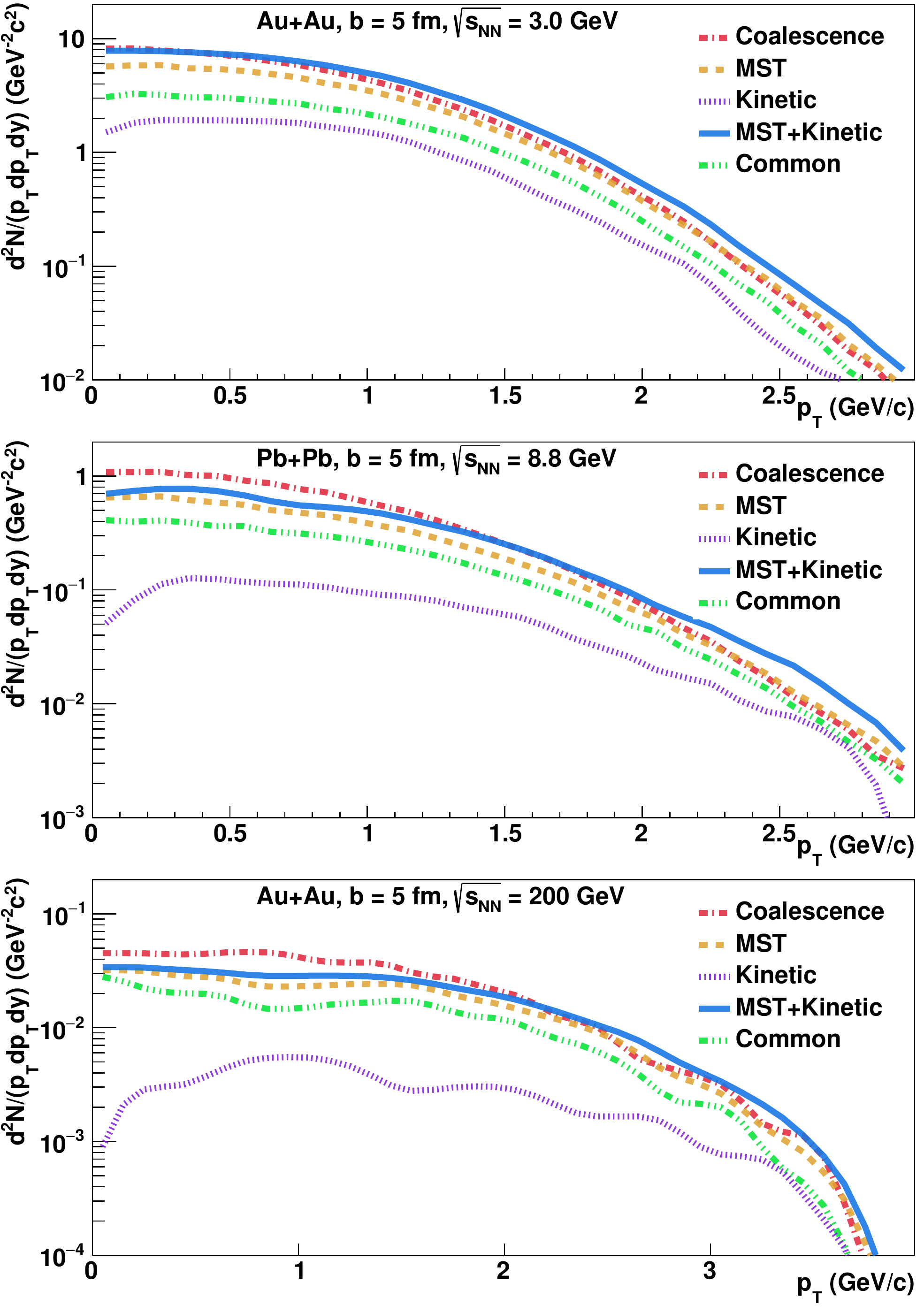}
    \caption{Transverse momentum distribution of deuterons produced by different scenarios at midrapidity ($|y| < 0.2$). The color coding is the same as in Fig \ref{fig:dndy}.}
    \label{fig:pt}
\end{figure}
\begin{figure}[t]
    \centering
        \includegraphics[width=0.45\textwidth]{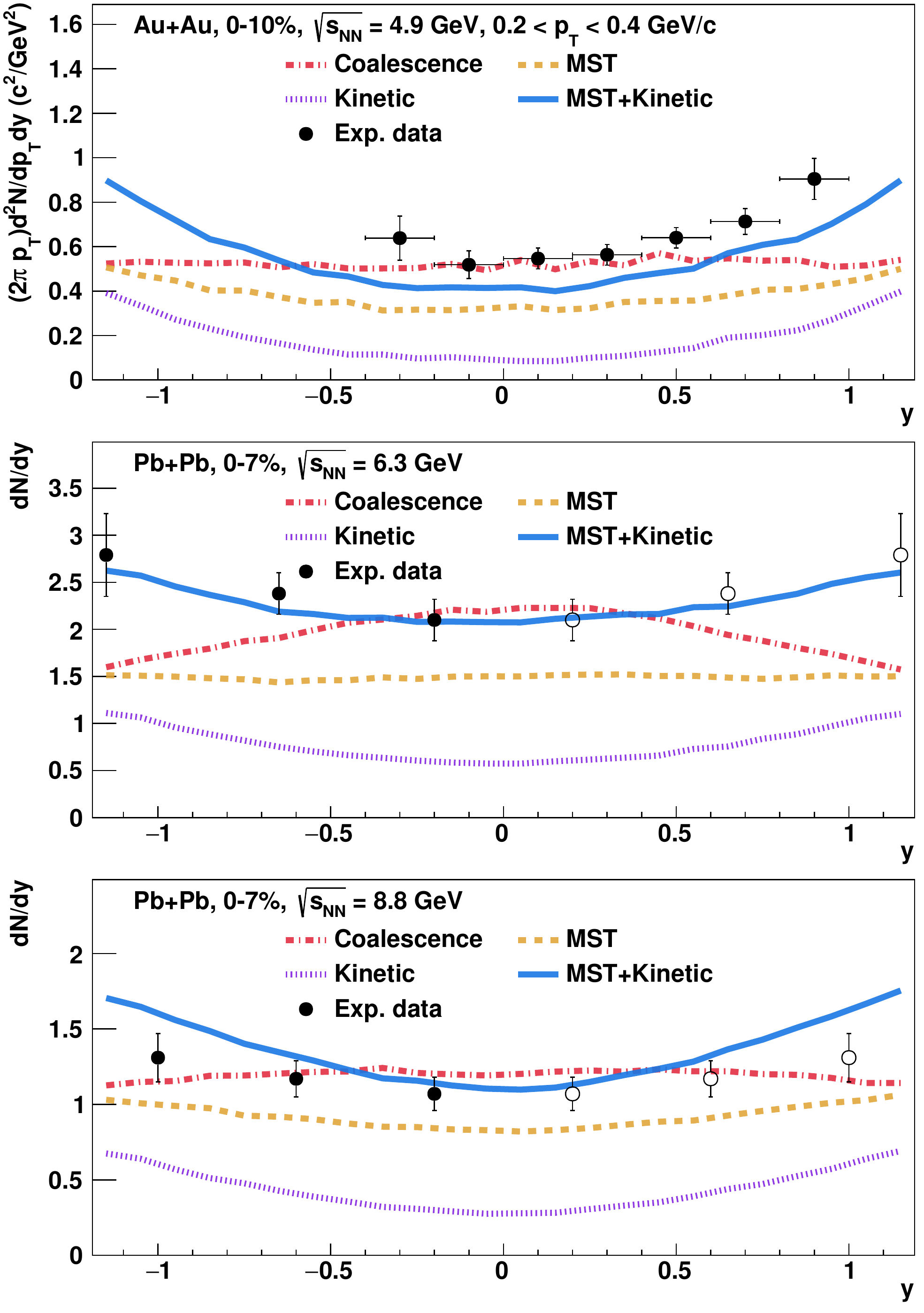} 
    \caption{The rapdity distribution of deuterons for the  10\% most central Au+Au collisions at $E_{Lab}=11$ AGeV ($\sqrt{s}=4.9$ GeV) measured by the E864 Collaboration in the interval $0.2 \le p_T \le 0.4$ GeV$/c$ \cite{Armstrong:2000gz} and for 7 \% most central Pb+Pb collisions from the NA49 Collaboration~ \cite{NA49:2016qvu} for $E_{Lab}=20$ and 40 AGeV 
    in comparison to the  PHQMD calculations for the different scenarios.
    The color coding is the same as in Fig \ref{fig:dndy}.}
    \label{fig:dndy_exp_comp}
\end{figure}

\begin{figure}[t]
    \centering
        \includegraphics[width=0.45\textwidth]{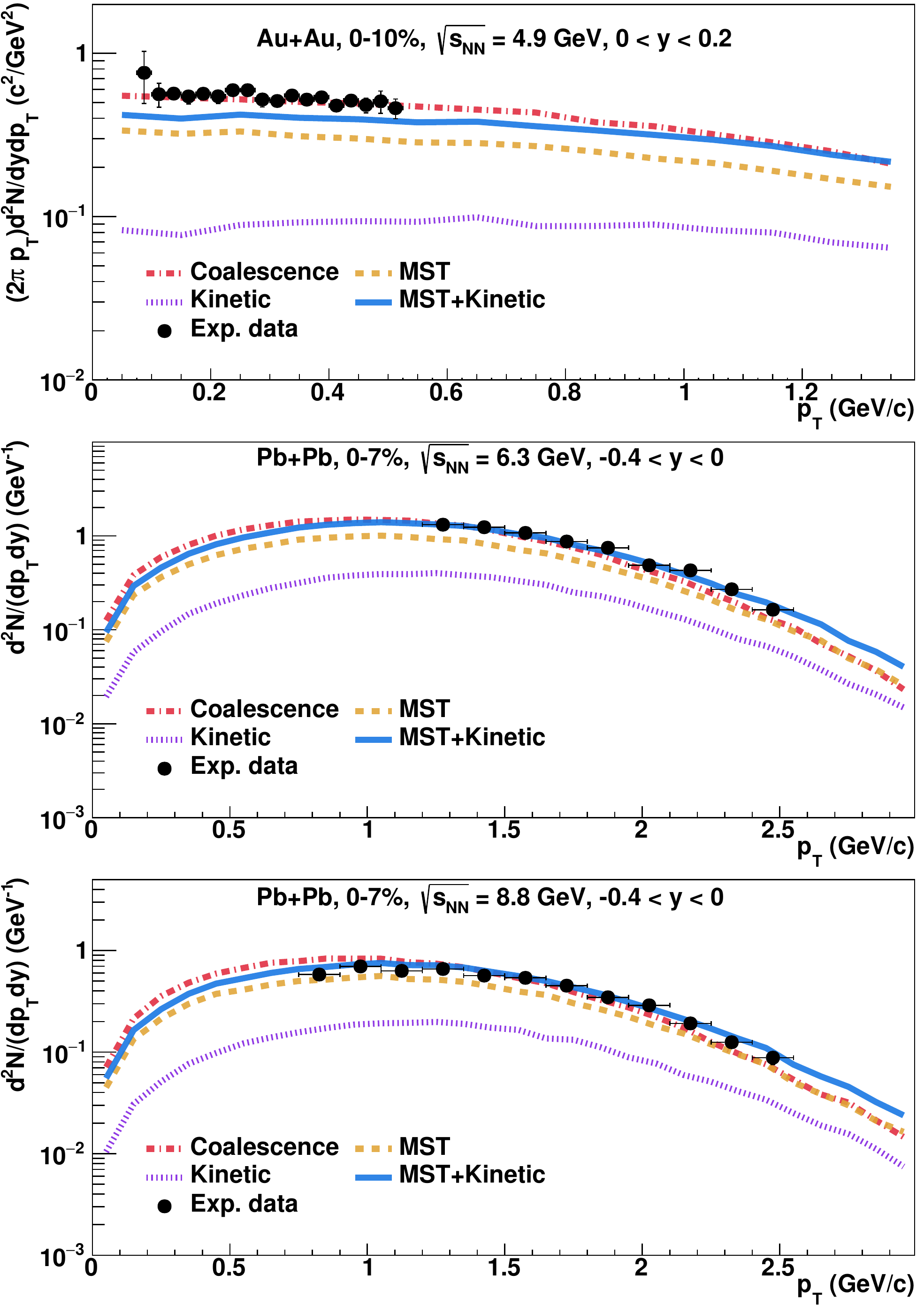} 
    \caption{The transverse momentum distributions of deuterons for the  10\% most central Au+Au collisions at $E_{Lab}=11$ AGeV ($\sqrt{s}=4.9$ GeV) in the rapidity interval $0 \le y \le 0.2$ from the E864 Collaboration~\cite{Armstrong:2000gz} and for 7 \% most central Pb+Pb collisions from the NA49 Collaboration~ \cite{NA49:2016qvu} for $E_{Lab}=20$ and 40 AGeV  in the rapidity interval $-0.4 \le y \le 0$ in comparison to the  PHQMD calculations for the different scenarios.
    The color coding is the same as in Fig \ref{fig:dndy}. }
    \label{fig:pt_exp_comp}
\end{figure}
From Fig.  \ref{fig:dndy} we see that for determining experimentally the production mechanism one has to measure experimentally the ratio $\frac{dN}{dy}(y=0)/\frac{dN}{dy}(y=0.6)$ for the lowest energy and the ratio $\frac{dN}{dy}(y=0)/\frac{dN}{dy}(y=1)$ for the higher energies with a high enough precision, which is, however experimentally achievable.  For this ratio we obtain 1.25 for $\sqrt{s}=3$ and 1.5 for the reactions at higher energies. This precision has almost been obtained in the NA49 experiment \cite{NA49:2016qvu} and will be achievable with more modern detectors. In \cite{NA49:2016qvu} the rapidity distribution of deuteron is convex around midrapidity.

Fig. \ref{fig:dndy_ptcut} shows the same rapidity distributions of deuterons as Fig. \ref{fig:dndy} but with the transverse momentum cut $p_{T} > 1.2$ GeV/c
in line with the present STAR acceptance. 
As seen in Fig. \ref{fig:dndy_ptcut}, a $p_T$ cut can change the form of the rapidity distributions at both, low and high energies. Thus, the measurement of low $p_{T}$ deuterons is necessary to identify the production mechanism.

Is this distinction also true for other observables? In Fig. \ref{fig:pt} we display the transverse momentum distribution of the deuterons for the same three energies.  The color coding is the same as in Fig. \ref{fig:dndy}. We observe that the distribution of `coalescence' and `potential' deuterons agree at large $p_T$ for the two lower energies, however, at small $p_T$ the spectra differ. This means that the clusters, which are produced by coalescence but which are not bound at the end (and therefore do not appear as `potential' deuterons), are concentrated at low $p_T$.  This can also be inferred from Figs.
\ref{fig:dndy} and \ref{fig:dndy_ptcut}. Thus, as said, one has to measure the $p_T$ spectra to low momentum in order to identify the deuteron production mechanism.

One may ask the question whether the presently available data in the interesting $\sqrt{s}$ region are already sufficient to identify the production mechanism for deuterons. To address this question we show in Figs. \ref{fig:dndy_exp_comp} and \ref{fig:pt_exp_comp} 
the PHQMD calculations for the different scenarios in comparison to  
the experimental rapidity and transverse momentum distributions of deuterons for the 10\% most central Au+Au collisions at $E_{Lab}=11$ AGeV ($\sqrt{s}=4.9$ GeV) from the E864 Collaboration~\cite{Armstrong:2000gz} and for the 7\% most central Pb+Pb collisions from the NA49 Collaboration~ \cite{NA49:2016qvu} for $E_{Lab}=20$ and 40 AGeV. 

We note that the experimental rapidity spectra are obtained by the integration of the $p_T$ spectra for different rapidity bins, which requires the extrapolation of the measured $p_T$ distributions to the low $p_T$ region, which is difficult to  measure in experiments. This extrapolation is usually done with a blast-wave fit. The slope of the $p_T$ spectra at low $p_T$ might differ from the blast-wave fit and, 
therefore, this extrapolation procedure introduces uncertainties in the integrated $dN/dy$ distributions. 

As follows from Fig. \ref{fig:pt_exp_comp}, in the measured $p_T$ range the NA49 transverse momentum spectra are nicely reproduced in the coalescence as well as in the MST + kinetic scenario. However, the coalescence spectra are slightly softer compared to the MST + kinetic spectra. The difference in the rapidity distributions in  Fig. \ref{fig:dndy_exp_comp} - obtained by the integrated $p_T$ yield - comes from the low $p_T$ region, where  experimental data are not available. 
Nevertheless, the shape of the extrapolated NA49 experimental $dN/dy$ spectra is more in favour to the MST + kinetic scenario for the both bombarding energies, 20 and 40 AGeV. 

In the E864 experiment the low $p_T$ spectra has been precisely measured but, as shown in Ref. \cite{Coci:2023daq}, the MST + kinetic scenario under-predicts the deuteron yield by about 15\% in all rapidity intervals. However, from the E864 data one may conclude with all caution that the presently available data favour aMST + kinetic scenario for deuteron production. More precise low $p_T$ deuteron data are certainly necessary to confirm this conclusion. It is also evident that a sufficient experimental precision is in reach. 

Furthermore, we have studied collective variables as the directed flow, $v_1$, and the elliptic flow, $v_2$, which are the first two coefficients of the Fourier series of the azimuthal angular distribution, as a function of the rapidity (integrated over the whole $p_T$ range). All three approaches provide rather similar results for these observables, which, therefore, may be not well suited to discriminate between the models for deuterons production since it would require high precision measurements of the $v_1$, $v_2$ flow coefficients, what is difficult to achieve experimentally.

\section{Conclusions}

We have investigated three approaches for deuteron production - coalescence, `potential and `kinetic' mechanisms, which have been advanced to explain the finite cluster yield at midrapidity, observed in ultra-relativistic heavy ion collisions. 

We have found that there are observables, which are sensitive to the deuteron production mechanism: the rapidity distribution has a different form and the transverse momentum distribution has a different slope at low $p_T$. These differences are large enough to be measurable and will allow, therefore, for discriminating between the different mechanisms for deuteron production when confronting these results with data. Knowing the mechanism we can also identify how deuterons survive the hot and dense medium created  at midrapidity of heavy-ion collisions and solve the `ice in the fire' puzzle.  \\
The analysis of the presently available data points tentatively to the MST + kinetic scenario but further experimental efforts are necessary to establish this mechanism.

\vspace*{3mm}
\textit{Acknowledgements:} The authors acknowledge inspiring discussions with M. Bleicher,  V. Kolesnikov, J. Steinheimer, Io. Vassiliev, V. Voronyuk and N. Xu as well as the support by the Deutsche Forschungsgemeinschaft  (DFG, German Research Foundation),  by the GSI-IN2P3 agreement under contract number 13-70. This study is part of a project that has received funding from the European Union's Horizon 2020 research and innovation program under grant agreement STRONG--2020 -- No 824093. V. Kireyeu acknowledge the support by the JINR young scientists grant 23-102-04.


\bibliography{biblio_PHQMD-d}

\end{document}